# Determination of apparent activating energy of solubilizing MLV-DPPC and MLV-egg-yolk lecithin with surfactants by rectangular optic diffusimetry

ŞTEFAN HOBAI
*Department of Biochemistry, University of Medicine and Pharmacy, Tg. Mureş, Romania*

*Correspondence address:*

Ştefan Hobai, Departament of Biochemistry, University of Medicine and Pharmacy, Târgu Mureş, Romania. Fax: 040-065-210407, E-mail: *fazy@netsoft.ro*

**Abstract**

It were determinated the apparent activating energy of dipalmitoylphosphatidylcholine (DPPC) multilamellar liposome (MLV) solubilizing with sodium deoxycholate (DOCNa) by turbidimetric measurements and the apparent activating energy of egg-yolk lecithin (EYL)-MLV solubilizing with different surfactants (DOCNa, TX-100, CTMB) by rectangular optic diffusimetric measurements. The apparent activating energy of MLV-DPPC solubilizing with DOCNa is $E_a$ = 127.7kJ/mol. The apparent activating energy of MLV-EYL solubilizing with cetyltrimethylamonium bromide (CTMB) is 22.3kJ/mol and with DOCNa is 47.3kJ/mol (average value) and with Triton X-100 is 99.3kJ/mol. On the basis of the above data it can be said that besides the utilisation of the liposomes in the modelling of cellular membranes, they can be used as experimental models, too, in order to investigate the solubilization capacity of surfactants.

**Introduction**

The interaction of various types of surfactants with lamellar phospholipid aggregates has been described in the literature [4,5,6]. Surfactants have been applied to study the size and structure of the formed mixed micelles [7,8,9] and the kinetics and the mechanism of membrane solubilization [2,3]. In some reviews is report the different mechanism of solubilization multilamellar and unilamellar liposomes [1,5].

**Materials**

Dipalmitoylphosphatidylcholine(di-$C_{16}$PC) (DPPC) was purchased from Sigma Chemical Co.
Egg yolk lecithine was purchased from Sigma Chemical Co. type X-E (cat.no.P5394/1996) purified by us by neutral alumina column chromatography, verified by thin layer chromatography (silicagel). The mobile phase was a mixture $CHCl_3:CH_3OH:H_2O$ (65:30:4,vol). The identification of the phospholipid was made with iodine and shows a single spot. Sodium deoxycholate (DOCNa) was purchased from Merck, the Triton X-100 (TX-100) from Sigma Chemical Co. and Cetyltrimethylammoniumbromid (CTMB) from Fluka AG. Tris(hydroxymethyl)aminomethane-HCl (Tris-Cl) was purchased from Austranal.

All chemicals and inorganic solvents were of analytical and of spectroscopic grades, respectively.

**Preparation of MLV-DPPC and MLV-egg-yolk lecithin (EYL) suspensions**

2.5 mg DPPC (SIGMA) disolved in 0.2 ml ethanol (redistilled on $P_2O_5$) or 0.2ml EYL (78% purity from SIGMA purified by us up to thin layer chromatographic single spot iodine) in chlorophorm:methanol (9:1,vol) solution which is in a glass test tube was evaporated under a nitrogen stream until it reached the consistence of a lipid film on the wall tube settled down. The film was purged with the gas for 30 minutes in order to eliminate the alcohol traces. The film hydration and bilayer suspension were made with sucrose (CHEMAPOL) solution of 16.6% at 0.6 mg/ml (lipid/solution) by vortexing 10 sec. above the chain-melting temperature ($T_m$ ~41,5°C).

**Methodology**

1. By using a spectrophotometer Specol and K-201 I titrated samples of 1.5ml solution of MLV-DPPC 0.6mg/ml (total 1.23µmol phospholipid) with DOCNa 20mM at a 4µl/sec titration rate up to a total addition of 16µmol surfactant. The titration was made at several temperatures between 16 and 22 °C.
2. By using a spectrophotometer Specol with FR optical diffusion system and photocells and voltage source type HQE 40 and UV lamp type OSRAM it was possible to determine the rectangular optic diffusion at λ=437nm of some MLV-egg-yolk lecithin suspension (1.4ml, 0.7µmol lipid/sample) in the course of the titration with the surfactant solution. The titration rate was 0.08µmol surfactant/sec. The tables below include the experimenal data concerning the total volume of titrant (expressed in µl), the total quantity of surfactant that was added (expressed in µmols) and modification of the percentage diffusion calculated by us being 1cm, as the deflexion unity of the recording pen (ΔDif%/cm). ΔDif%/µmol surfactant is the variation of the unity of percentage diffusion, calculated for the 1µmol surfactant addition.

**Results and discussion**

1. It is known that the titration curves in the transmission system of SUV-egg-yolk lecithin have a sigmoidal form, and in the case of MLV-DPPC they are straight.

Table 1 includes the experimental data obtained by turbidimetric titrations.

**Table 1**. *Variation speed of optical transmission percentage (Tr%) during the titration of the MLV-DPPC solutions with DOCNa at different temperatures.*

| Nr. samples | t (°C) | T (K) | $10^3/T$ | ΔTr % | ln(ΔTr %) |
|---|---|---|---|---|---|
| 1 | 16 | 289 | 3.458 | 3.36 | 1.21 |
| 2 | 17.1 | 290.1 | 3.445 | 5.08 | 1.62 |
| 3 | 18.4 | 291.4 | 3.430 | 6.35 | 1.85 |
| 4 | 20.2 | 293.2 | 3.409 | 6.98 | 1.94 |
| 5 | 21 | 294 | 3.400 | 9.44 | 2.24 |
| 6 | 22 | 295 | 3.389 | 10.95 | 2.39 |

Table 2 presents the necessary data for the calculation of the regression line of dependence ln(ΔTr%) ~ f($10^3$/T) necessary to determine the activating energy of MLV-DPPC solubilization with DOCNa.

**Table 2**. *Statistical data of titration with DOCNa 20mM with a rate 4μl/sec of 1.5ml MLV-DPPC (1.23μmol of phospholipid) at different temperatures. $X_n=10^3/T$; $Y_n= ln(ΔTr\%)$. T-absolute temperature (K); ΔTr%- variation of optical transmission percentage. The horizontal bar indicates the mean values. Σ is the sum of the individual values.*

| n | $X_n$ | $Y_n$ | $X_n - \bar{X}$ | $Y_n - \bar{Y}$ | $(X_n - \bar{X})(Y_n - \bar{Y})$ | $(X_n - \bar{X})^2$ | $(Y_n - \bar{Y})^2$ |
|---|---|---|---|---|---|---|---|
| 1 | 3.389 | 2.39 | -0.032 | 0.515 | -0.01648 | 0.001024 | 0.265 |
| 2 | 3.400 | 2.24 | -0.021 | 0.365 | -0.007665 | 0.000441 | 0.1332 |
| 3 | 3.409 | 1.94 | -0.012 | 0.065 | -0.00078 | 0.000144 | 0.00422 |
| 4 | 3.430 | 1.85 | 0.009 | -0.025 | -0.000225 | 0.000081 | 0.000625 |
| 5 | 3.445 | 1.62 | 0.024 | -0.255 | -0.00612 | 0.000576 | 0.065 |
| 6 | 3.458 | 1.21 | 0.037 | -0.665 | -0.0246 | 0.001369 | 0.4422 |
| Σ | 20.531 | 11.25 | - | - | -0.055875 | 0.003635 | 0.91027 |
| mean | 3.421 | 1.875 | | | | | |

**Figure 1**. *Arrhenius logarithmic representation of change of speed of optical transmission percentage (logarithmic) of sample MLV-DPPC during the titration with DOCNa at different temperatures. The straight line in the middle of experimental points was obtained by linear regression. Statistical data appear in Table 2.*

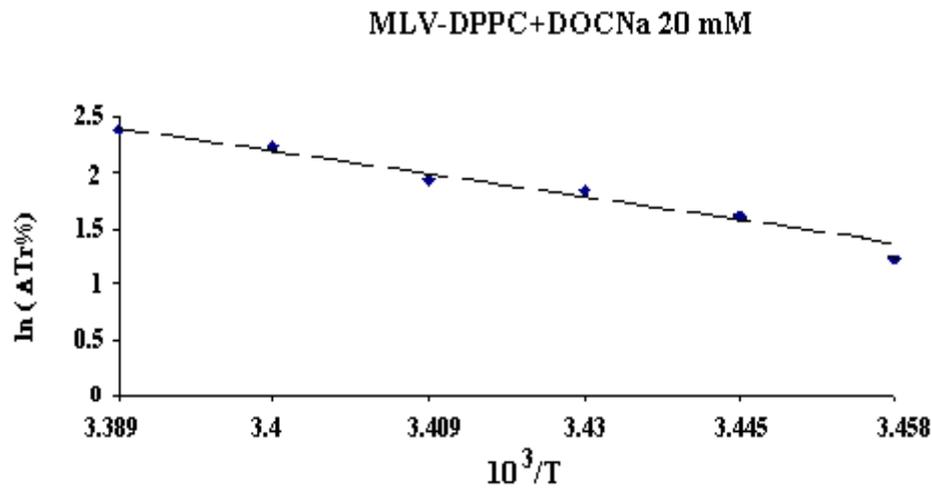

The correlation coefficient (r) was calculated based on data in Table 2.

$$r = \frac{\sum_{1}^{6}(X_n - \bar{X})(Y_n - \bar{Y})}{\sqrt{\sum_{1}^{6}(X_n - \bar{X})^2 \sum_{1}^{6}(Y_n - \bar{Y})^2}} = \frac{-0.055875}{\sqrt{0.003635 \cdot 0.91027}} = \frac{-0.055875}{0.05752} = -0.9714 \quad (1)$$

The correlation coefficient value indicates a good linear correlation between the two experimental data $X_n$ and $Y_n$. The calculus of the straight line of linear regression produced the following data:

$$a = \frac{\sum_{1}^{6}(X_n - \bar{X})(Y_n - \bar{Y})}{\sum_{1}^{6}(X_n - \bar{X})^2} = \frac{-0.055875}{0.003635} = -15.37 \quad (2)$$

$$b = \bar{Y} - \bar{X} \cdot a \approx 1.875 - 3.421(-15.37) = 54.45 \quad (3)$$

The expression of straight line regression is:

$$Y = -15.37X + 54.45 \quad (4)$$

Straight line regressin slope is equal to $E_a / R$ ratio, so the apparent activating energy of solubilization is $E_a = R \cdot 15.37 = 127.7 \text{kJ/mol}$.

2. Determination of apparent activating energy of solubilizing MLV-EYL with surfactants (DOCNa, TX-100, CTMB) by rectangular optic diffusimetry was carried out by the same procedure as above.

**Table 3**. *Titrant: DOCNa 20mM; t= 17.5°C; ΔDif.% /cm = 5.5*

|   | μl DOCNa | μmol DOCNa | ΔDif%/μmol DOCNa |
|---|---|---|---|
| 1 | - | - | - |
| 2 | 12 | 0.24 | 22.92 |
| 3 | 8 | 0.16 | 34.37 |
| 4 | 8 | 0.16 | 34.37 |
| 5 | 24 | 0.48 | 11.46 |
| 6 | 16 | 0.32 | 17.19 |
| 7 | 10.4 | 0.2 | 27.50 |
| 8 | 29.6 | 0.6 | 9.16 |
| 9 | 28 | 0.56 | 9.82 |
| 10 | 28 | 0.56 | 9.82 |
| Mean ΔDif%/μmol DOCNa = | | | 19.7 |
| t =22.5°C; ΔDif.%/cm =5.68 | | | |
| 1 | - | - | - |
| 2 | 8 | 0.16 | 35.5 |
| 3 | 20 | 0.4 | 14.2 |
| 4 | 8 | 0.16 | 35.5 |
| 5 | 4 | 0.08 | 71 |
| 6 | 8 | 0.16 | 35.5 |
| 7 | 28 | 0.56 | 10.1 |
| 8 | 16 | 0.32 | 17.7 |
| 9 | 24 | 0.48 | 11.8 |
| 10 | 20 | 0.4 | 14.2 |
| Mean ΔDif%/μmol DOCNa = | | | 27.3 |

For the calculus of the apparent activating energy, $E_a$, of solubilizing MLV-egg-yolk we use the following formulae:

$$\frac{\ln(\Delta Dif\%/\mu mol)_2 - \ln(\Delta Dif\%/\mu mol)_1}{1000/T_2 - 1000/T_1} = \frac{3.307 - 2.981}{3.384 - 3.442} = 5.62$$

$E_a = 5.62 \cdot R = 46.7$ kJ/mol

**Table 4**. *Titrant: DOCNa 20mM; t= 17.5°C; ΔDif.%/cm = 6.58*

|   | μl DOCNa | μmols DOCNa | ΔDif%/μmol DOCNa |
|---|---|---|---|
| 1 | - | - | - |
| 2 | 12 | 0.24 | 27.4 |
| 3 | 24 | 0.48 | 13.7 |
| 4 | 21.6 | 0.43 | 15.3 |
| 5 | 26.4 | 0.53 | 12.4 |
| 6 | 36 | 0.72 | 9.14 |
| 7 | 42 | 0.84 | 7.83 |
| 8 | 42 | 0.84 | 7.83 |
| 9 | 132 | 2.64 | 2.49 |
| Mean ΔDif%/μmol DOCNa = | | | 12 |
| t = 21°C; ΔDif.%/cm=5.26 | | | |
| 1 | - | - | - |
| 2 | 6 | 0.12 | 43.8 |
| 3 | 27.6 | 0.55 | 9.56 |
| 4 | 20.4 | 0.41 | 12.8 |
| 5 | 15.6 | 0.31 | 16.97 |
| 6 | 32.4 | 0.65 | 8.1 |
| 7 | 24 | 0.48 | 10.9 |
| 8 | 60 | 1.2 | 4.38 |
| Mean ΔDif%/μmol DOCNa = | | | 15.2 |

$$\frac{\ln(\Delta Dif\%/\mu mol)_2 - \ln(\Delta Dif\%/\mu mol)_1}{1000/T_2 - 1000/T_1} = \frac{2.722 - 2.485}{3.401 - 3.442} = 5.78$$

$E_a = 5.78 \cdot R = 48$ kJ/mol

It can be observed that the apparent activating energy of solubilizing MLV-egg-yolk with DOCNa, determined in two different experiments (made on different days), had really close values (46.7 kJ/mol and 48kJ/mol, respectively) proving a good reproductibility of the determination.

**Table 5**. *Titrant: Triton X-100 20mM; t= 17.5°C; ΔDif.%/cm = 5.34*

|   | μl T X-100 | μmols T X-100 | ΔDif%/μmol T X-100 |
|---|---|---|---|
| 1 | - | - | - |
| 2 | 12 | 0.24 | 22.2 |
| 3 | 12 | 0.24 | 22.2 |
| 4 | 12 | 0.24 | 22.2 |
| 5 | 12 | 0.24 | 22.2 |
| 6 | 12 | 0.24 | 22.2 |
| 7 | 96 | 0.19 | 28.1 |
| 8 | 12 | 0.24 | 22.2 |
| 9 | 26.4 | 0.53 | 10.1 |
| 10 | 36 | 0.72 | 7.42 |
| Mean ΔDif%/μmol TX-100 = | | | 19.87 |
| t = 21°C; ΔDif.%/cm = 6.06 | | | |
| 1 | - | - | - |
| 2 | 6 | 0.12 | 50.5 |
| 3 | 8.4 | 0.17 | 35.6 |
| 4 | 7.2 | 0.14 | 43.3 |
| 5 | 12 | 0.24 | 25.25 |
| 6 | 4.8 | 0.1 | 60.6 |
| 7 | 19.2 | 0.38 | 15.9 |
| 8 | 8.4 | 0.17 | 35.6 |
| 9 | 18 | 0.36 | 16.8 |
| 10 | 36 | 0.72 | 8.4 |
| Media ΔDif%/μmol TX-100 = | | | 32.44 |

$$\frac{\ln(\Delta Dif\%/\mu mol)_2 - \ln(\Delta Dif\%/\mu mol)_1}{1000/T_2 - 1000/T_1} = \frac{3.479 - 2.989}{3.401 - 3.442} = 11.95$$

$E_a$= 11.95·R= 99.314 kJ/mol

**Table 6**. *ΔDif%/μmol is calculated on the basis of knowledge of the variation of percentage diffusion determined, by adding 8μmols surfactant to the sample.*

| Nr | (°C) | 1000/T (grad$^{-1}$) | ΔDif%/μmol | ln(ΔDif%/μmol) CTMB |
|---|---|---|---|---|
| 1 | 18.4 | 3.431 | 8.89 | 2.185 |
| 2 | 20.5 | 3.407 | 9.517 | 2.253 |
| 3 | 22.5 | 3.384 | 10.08 | 2.311 |

**Figure 2**. *The Arrhenius representation of the logarithm of the variation of optical diffusion percentage to 1μmol CTMB at different temperatures. The straight lune was obtained by linear regression.*

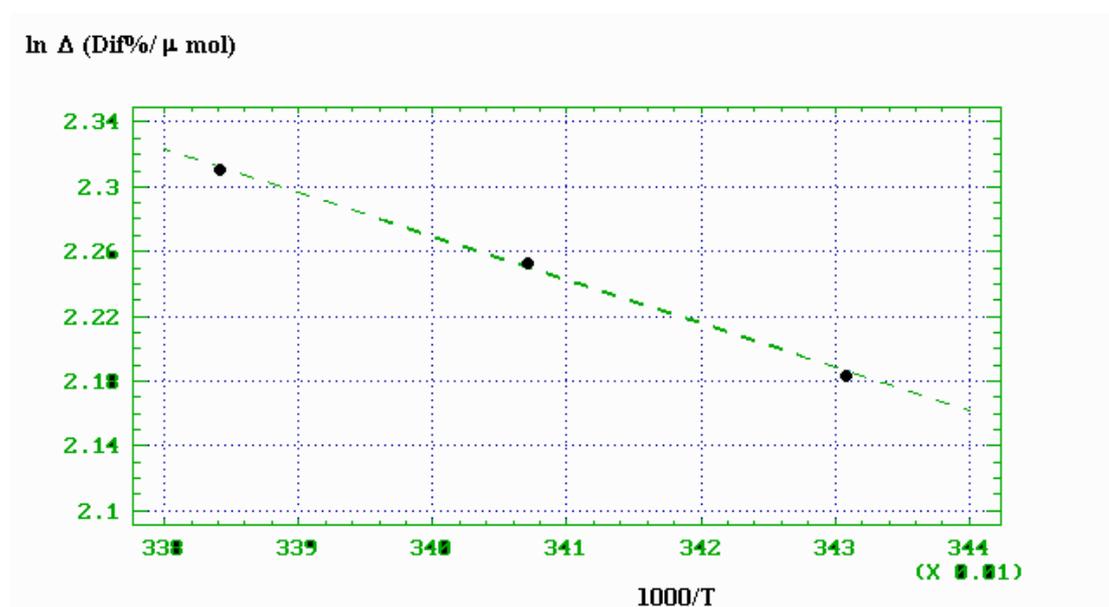

$$\frac{\Delta \ln(\Delta Dif\%/\mu mol)}{1000/T_2 - 1000/T_1} = \frac{2.311 - 2.185}{3.384 - 3.431} = 2.68$$

$E_a = 2.68 \cdot R = 22.3$ kJ/mol

In the Table 7 the apparent activating energies for the solubilizations with the three surfactants are presented.

**Table 7**. *The apparent activating energy of solubilizing MLV-EYL with surfactants.*

| Surfactants | $E_a$ (kJ/mol) |
|---|---|
| CTMB | 22.3 |
| DOCNa | 48 and 46.7 |
| Triton X-100 | 99.3 |

It is observed that the solubilization powers expressed in kinetic terms, in the condition of titrating at a rate of 0.08μmols surfactant/sec, for the three surfactants, can be arranged in the order: CTMB<DOCNa<TX-100.

**Glossary**

CMC- critical micellar concetration

CTMB – cetyltrimethylammonium bromide

Dif%- percentage optical diffusion

DOCNa-sodium deoxycholate

DPPC-dipalmitoylphosphatidylcholine

EYL – egg-yolk lecithin

$E_a$ – apparent activating energy

MLV- multilamellar vesicles

r- correlation coefficient

SUV- small unilamellar vesicles

Tr%- percentage optical transmission

TX-100- Triton X-100, polyoxyethylene-p-t-octilfenol